\documentclass[apj,twocolumn]{openjournal}
\pdfoutput=1
\usepackage{graphicx}
\usepackage{enumerate}
\usepackage{amssymb, amsmath, amsfonts}
\usepackage{hyperref}
\usepackage{ulem}
\usepackage{url}
\usepackage{xspace}
\usepackage{xcolor}

\defcitealias{Torres_1999}{T99}

\usepackage{orcidlink}

\begin{document}

\title{Closed-Form Statistical Relations Between Projected Separation, Semimajor Axis, Companion Mass, and Host Acceleration}

\author{\vspace{-30pt}Timothy D.~Brandt\,\orcidlink{0000-0003-2630-8073}$^{1}$}
\affiliation{$^1$Space Telescope Science Institute, 3700 San Martin Drive, Baltimore, MD, 21218, USA}
\email{Corresponding author: tbrandt@stsci.edu}

\begin{abstract}
I derive the statistical relationship between a radial velocity or astrometric acceleration (a trend), a companion's mass, and the projected separation of the companion.  These relationships, expressed as probability density functions, are analytic and independent of all Keplerian orbital elements so long as orbits are randomly oriented in space.  I also derive a closed-form expression for the probability distribution of the ratio of the projected separation to the semimajor axis at fixed eccentricity.  This expression can be numerically integrated over eccentricity for an arbitrary distribution of eccentricities.  I verify my results with empirical comparisons to equivalent but more complex expressions in the literature based on the equations of Keplerian orbits.  The closed-formed expressions derived here would be especially useful for any calculation that requires derivatives, e.g., Hamiltonian Monte Carlo.  I also provide a Jupyter notebook including all figures and calculations at \url{https://github.com/
t-brandt/analytic-orbit-calculations}. 
\end{abstract}

\maketitle

\section{Introduction}

Stars with widely separated companions undergo orbital motion, but the periods involved can be much longer than available observational baselines.  In this case we see only a short orbital arc.  For the host star, this may appear as an acceleration.  A radial acceleration is observed in radial velocities \citep[RVs,][]{Griffin+Gunn+Zimmerman+Griffin_1988,Mermilliod+Rosvick+Duquennoy+Mayor_1992}; a sky-plane acceleration may be detected using absolute astrometry \citep{Bessel_1844}.  These accelerations, especially in RV, are also known as trends.  We may also observe the companion itself, enabling the measurement of a projected separation \citep{Crepp+Johnson+Howard+etal_2012,Crepp+Johnson+Howard+etal_2013,Bowler+Cochran+Endl+etal_2021}, and in some cases the relative sky-plane motion of the primary and secondary may be available.  This is increasingly the case with data from the Gaia spacecraft \citep{Gaia_General_2016}.

\cite{Torres_1999}, hereafter \citetalias{Torres_1999}, was the first to systematically explore the constraints on companion mass that are provided by short orbital arcs.  Using Monte Carlo, they computed the probability distributions of companion mass as a function of projected separation for a measured acceleration of the host star in either RV or astrometry.  \citetalias{Torres_1999} also computed probability distributions of the ratio of projected separation to semimajor axis given a few different underlying distributions of eccentricity.  These distributions were likewise calculated using Monte Carlo.  \cite{Savransky+Cady+Kasdin_2011} developed an analytic framework for computing distributions of true and projected separations, while \cite{Garrett+Savransky+Macintosh_2017} applied the results of \cite{Savransky+Cady+Kasdin_2011} to direct imaging surveys in the case of zero eccentricity.

In this paper I derive closed-form expressions equivalent to those computed in \citetalias{Torres_1999}.  For RV or astrometric accelerations, these expressions may be obtained using Newton's Law of Gravity and assuming the separation vector to be randomly oriented, with no dependence on the orbital elements.  Section \ref{sec:analytic} derives analytic expressions for the probability distributions of mass as a function of measured acceleration and projected separation in both the astrometric and the RV case.  Sections \ref{sec:equivalence_1} and \ref{sec:equivalence_2} demonstrate the equivalence of these expressions to those of \citetalias{Torres_1999}.  Section \ref{sec:projsep_semimajoraxis} derives a closed-form expression for the probability distribution of the ratio of the projected separation to semimajor axis at fixed eccentricity.  This must, in general, be integrated numerically over a distribution of eccentricities.  I conclude with Section \ref{sec:conclusions}.

\section{Analytic Expressions for Trends} \label{sec:analytic}

I first derive analytic expressions for the relation between the mass of a companion to a star, the star's observed acceleration, and the companion's projected separation.  Denoting the mass of a faint companion tugging a star as $M_B$, the RV and astrometric accelerations may be derived from Newton's Law of Gravity as
\begin{equation}
    a_{\rm RV} = \frac{d{\rm RV}}{dt} = \frac{GM_B}{r^2} \cos \varphi
    \label{eq:rvtrend}
\end{equation}
and 
\begin{equation}
    a_{\rm ast} = \frac{G M_B}{r^2} \sin \varphi \label{eq:accel_astrometric}
\end{equation}
where $\varphi$ is the angle between the line-of-sight and the separation vector ${\bf r}$ between the star and its companion.  A value of $\varphi=0$ indicates a star with a companion displaced along the line-of-sight.  This is geometrically less likely than a separation vector in the sky plane (with $\varphi=\pi/2$).  Throughout this paper I will consider only the absolute values of accelerations, of $\sin \varphi$, and of $\cos \varphi$, so that it is sufficient to take $\varphi \in [0, \pi/2]$.  Equation \eqref{eq:accel_astrometric} assumes that only the acceleration of the primary star is measured, i.e., that there is no light from the companion contaminating the measurement.  The projected separation, the product of the distance $D$ from Earth to the star and the angular separation in radians $\rho$, is
\begin{equation}
    D\rho = r \sin \varphi.
    \label{eq:projsep}
\end{equation}

I treat the astrometric case first.  I combine Equations \eqref{eq:accel_astrometric} and \eqref{eq:projsep} to get $M_B$ as a function of astrometric acceleration, projected separation $D \rho$, and $\varphi$, as
\begin{equation}
    M_B = \frac{1}{G} a_{\rm ast} (D\rho)^2 \Phi_{\rm ast}(\varphi)
    \label{eq:basic_eq}
\end{equation}
with
\begin{equation}
    \Phi_{\rm ast} = \frac{1}{\sin^3 \varphi}.
    \label{eq:Phi_ast}
\end{equation}
The quantity $\Phi_{\rm ast}$ increases monotonically with $\cos \varphi$ from $\cos \varphi = 0$ to $\cos \varphi = 1$ when $\varphi \in (0, \pi/2]$.  Assuming the orientation of the separation vector to be random on the sphere, the cumulative distribution function, or cdf, for $\cos \varphi$ may be written as ${\rm cdf}(\cos\varphi) = \cos\varphi$ for $\varphi \in [0, \pi/2]$.  These two facts allow me to write
\begin{align}
    {\rm cdf}(\Phi_{\rm ast}) = \cos \left(\varphi(\Phi_{\rm ast})\right) . 
    \label{eq:cdf_ast_part1}
\end{align}
This yields
\begin{align}
    {\rm cdf}(\Phi_{\rm ast}) = \sqrt{1 - \Phi_{\rm ast}^{-2/3}} ,
\end{align}
which has a median value of $(4/3)^{3/2} \approx 1.54$.  I can differentiate to obtain the probability density,
\begin{align}
    \frac{dp}{d\Phi_{\rm ast}} = \frac{1}{3\sqrt{\Phi_{\rm ast}^{10/3} - \Phi_{\rm ast}^{8/3}}} . \label{eq:Phi_ast_dist}
\end{align}
The domain of $\Phi_{\rm ast}$ runs from one to infinity.  

I now turn to the RV case, 
\begin{equation}
    M_B = \frac{1}{G} a_{\rm RV} (D\rho)^2 \Phi_{\rm RV}(\varphi)
\end{equation}
with
\begin{align}
    \Phi_{\rm RV} &= \frac{1}{\cos \varphi \sin^2 \varphi} \nonumber \\
    &= \frac{1}{\cos \varphi - \cos^3 \varphi} . \label{eq:Phi_RV}
\end{align}
Unfortunately, $\Phi_{\rm RV}(\cos\varphi)$ has a cubic denominator, and its inverse is multiply valued.  The two (real) solutions of 
\begin{equation}
    \cos \varphi - \cos^3 \varphi = \frac{1}{\Phi_{\rm RV} }
\end{equation}
that apply between $\varphi=0$ and $\varphi=\pi/2$ are
\begin{align}
    \cos\varphi_1 &= \frac{2}{\sqrt{3}} \cos \theta
    \label{eq:root1}
\end{align}
and
\begin{equation}
    \cos\varphi_2 = \sin \theta - \frac{1}{\sqrt{3}} \cos \theta
    \label{eq:root2}
\end{equation}
with
\begin{equation}
    \theta = \frac{1}{3} \cos^{-1} \left( -\frac{\sqrt{27}}{2\Phi_{\rm RV}} \right) .
    \label{eq:thetadef}
\end{equation}
Each of the equations for $\varphi_1$ and $\varphi_2$ applies over a different range of $\varphi$ values: Equation \eqref{eq:root1} applies between $\varphi = 0$ and $\varphi = \cos^{-1} \frac{1}{\sqrt{3}}$, and Equation \eqref{eq:root2} applies between $\varphi = \cos^{-1} \frac{1}{\sqrt{3}}$ and $\varphi = \pi/2$. Each interval in $\varphi$ corresponds to $\theta \in [\pi/6, \pi/3]$.  The function $\Phi_{\rm RV}(\varphi)$ has a minimum at $\cos \varphi = \frac{1}{\sqrt{3}}$.  

I again begin with the cdf.  The quantity $\Phi_{\rm RV}$ is monotonically increasing with $\cos\varphi$ from $\cos\varphi = \frac{1}{\sqrt{3}}$ to $\cos\varphi = 1$ (i.e.~for $\varphi_1$), and monotonically decreasing with $\cos\varphi$ from $\cos\varphi = 0$ to $\cos\varphi = \frac{1}{\sqrt{3}}$ (i.e.~for $\varphi_2$).  Using ${\rm cdf}(\cos\varphi) = \cos\varphi$, I may write the cdf as the sum of contributions from the two disjoint intervals in $\varphi$,
\begin{align}
    {\rm cdf}(\Phi_{\rm RV}) = &\cos \left(\varphi_1(\Phi_{\rm RV})\right) - \cos \left(\varphi_1(\Phi_{\rm RV, min}) \right) \nonumber \\
    & - \left( \cos \left(\varphi_2(\Phi_{\rm RV})\right) - \cos \left(\varphi_2(\Phi_{\rm RV, min}) \right) \right) .
    \label{eq:cdf_part1}
\end{align}
Using $\varphi_1(\Phi_{\rm RV, min}) = \varphi_2(\Phi_{\rm RV, min}) = \frac{1}{\sqrt{3}}$, this yields
\begin{align}
    {\rm cdf}(\Phi_{\rm RV}) &= \sqrt{3} \cos \theta - \sin \theta 
    \label{eq:PhiRV_cdf}
\end{align}
with $\theta$ from Equation \eqref{eq:thetadef}.  
The probability density may be obtained by differentiating Equation \eqref{eq:PhiRV_cdf}.  After some algebra, I have
\begin{align}
    \frac{dp}{d\Phi_{\rm RV}} &= \frac{3 \sin \theta + \sqrt{3}\cos \theta}{\sqrt{4 \Phi_{\rm RV}^4 - 27 \Phi_{\rm RV}^2}}.
    \label{eq:Phi_RV_final}
\end{align}
The limits on $\Phi_{\rm RV}$ run from  $\sqrt{27}/2 \approx 2.6$ \citep{Liu+Fischer+Graham+etal_2002} to infinity.  Neither $dp/d\Phi_{\rm RV}$ nor $dp/d\Phi_{\rm ast}$ depend on the orbital elements, so long as the orientation of the separation vector between the star and its companion is random.  The orbital elements are determined by the relationship of the position and velocity, but the acceleration of the host star is determined by the relative position alone.

\section{Equivalence to a Keplerian Derivation in RV} \label{sec:equivalence_1}

In this section I demonstrate the equivalence of Equation \eqref{eq:Phi_RV} to the corresponding expression of \citetalias{Torres_1999}.  The expressions of \citetalias{Torres_1999} are derived from the equations describing a Keplerian orbit, and the resulting probability distributions must be evaluated by Monte Carlo.  I derive the explicit equivalence for the RV case up to the definition of $\Phi_{\rm RV}$.  I defer a full numerical validation of both the RV and astrometric results to the following section.  From \citetalias{Torres_1999}, I have
\begin{equation}
    \frac{M_B}{M_\odot} = 5.341 \times 10^{-6} \left( \frac{D\rho}{\rm AU} \right)^2 \bigg| \frac{d{\rm RV}}{dt} \bigg| \Phi_{\rm RV}\left[i,e,\omega,\phi\right]
\end{equation}
with
\begin{align}
    \Phi_{\rm RV}\left[i,e,\omega,\phi\right] = (&1 - e)(1 + \cos E) \nonumber\\
    &\!\!\!\!\times \big[(1 - e \cos E) \left(1 - \sin^2 (\nu + \omega) \sin^2 i \right) \nonumber\\ &\qquad \times \sin (\nu + \omega)(1 + \cos \nu) \sin i \big]^{-1}.
    \label{eq:phasefunction}
\end{align}

In Equation \eqref{eq:phasefunction}, $e$ is the eccentricity, $E$ is the eccentric anomaly, $\nu$ is the true anomaly, $i$ is the inclination, $\omega$ is the argument of periastron, and $\phi$ is the phase (the mean anomaly), which is implicit in both the eccentric anomaly and the true anomaly.  

The equation simplifies considerably if I expand the term $1 + \cos \nu$.  The cosine of the true anomaly $\cos \nu$ is related to the eccentric anomaly $E$ by
\begin{align}
    \cos \nu &= \frac{\cos E - e}{1 - e\cos E} .
\end{align}
The phase function calls for $1 + \cos \nu$, which, after factoring, may be written as
\begin{align}
    1 + \cos \nu &= \frac{(1 - e)(1 + \cos E)}{1 - e \cos E}.
\end{align}
Plugging this into Equation \eqref{eq:phasefunction} gives
\begin{equation}
    \Phi_{\rm RV}\left[i,e,\omega,\phi\right] = \frac{1}{\sin(\omega + \nu) \sin i  - \sin^3 (\omega + \nu) \sin^3 i}. \label{eq:phasefunction_simplified}
\end{equation}
In Equation \eqref{eq:phasefunction_simplified}, the only remaining dependence on eccentricity is that implicit in the true anomaly $\nu$.  I define the argument of latitude $u = \omega + \nu$, writing Equation \eqref{eq:phasefunction_simplified} as
\begin{equation}
    \Phi_{\rm RV}\left[i,u \right] = \frac{1}{\sin u \sin i  - \sin^3 u \sin^3 i} .\label{eq:Phi_Torres_simplified}
\end{equation}
The geometry of the orbit gives $\sin i \sin u = \cos \varphi$, where $\varphi$ is the angle between the instantaneous separation vector of the companion and the line-of-sight.  Equation \eqref{eq:Phi_Torres_simplified} is thus identical to Equation \eqref{eq:Phi_RV}: so long as the instantaneous separation vector is randomly oriented in space, $\Phi_{\rm RV}$ has no dependence on eccentricity.

I may also derive the equivalence of Equations \eqref{eq:Phi_Torres_simplified} and \eqref{eq:Phi_RV} assuming the standard prior on $\omega$, i.e., one that is uniform in the argument of periastron $\omega$.  This is a consequence of assuming the orientation of the angular momentum vector to be random on the unit sphere.  If $\omega$ is uniformly distributed, then $u = \nu + \omega$ is also uniformly distributed for any well-defined distribution of $\nu$.  This is because phase-wrapping the sum of any quantity and a uniformly distributed variable at the limits of the uniform distribution gives a uniformly distributed result.  

Assuming a uniform prior on $u$ and a $\sin i$ prior on inclination, the distribution of $z = \sin u \sin i$ is, in fact, uniform: it is the same as the distribution of $\cos i$ under random orbital orientations.  This can be proved using the conditional probability integral for $z=xy$:
\begin{align}
    \frac{dp}{dz} = \int \left( \frac{dp}{dx} \right) \left( \frac{dx}{|x|} \right) \left( \frac{dp}{dy}\bigg|_{y=z/x} \right) . \label{eq:conditional_probability}
\end{align}
I take $x = \sin u$ and $y=\sin i$ and limit myself to the case where $\sin u$ and $\sin i$ are positive for simplicity.  I have
\begin{align}
    \frac{dp}{dz} = \int_{z}^1 \frac{1}{2\pi} \frac{dx}{x\sqrt{1-x^2}} \frac{z/x}{\sqrt{1-(z/x)^2}} = \frac{1}{4}.
\end{align}
Another factor of 2 comes from both sines being negative, while the situation is symmetric for $z<0$.  I thus have $dp/dz=\frac{1}{2} = {\rm constant}$ for $z \in [-1, 1]$, as I wished to show.  This approach also demonstrates that Equation \eqref{eq:Phi_Torres_simplified} is equivalent to Equation \eqref{eq:Phi_RV}, with no remaining dependence on any orbital elements. 

\section{Empirical Verification of Trend Results} \label{sec:equivalence_2}

I now empirically demonstrate the equivalence of my expressions for $\Phi_{\rm RV}$ and $\Phi_{\rm ast}$ in Equations \eqref{eq:Phi_RV_final} and \eqref{eq:Phi_ast_dist} to the corresponding equations in \citetalias{Torres_1999}, their Equations (6), (7), and (15), and (16).  For this exercise, I adopt a uniform distribution of eccentricities between 0 and 0.8, uniform distributions of the orbital parameters $\omega$, $\Omega$ (position angle of the ascending node), and orbital phase, and a geometric prior on inclination.  The actual prior on eccentricity does not matter; my results would be the same for another distribution.  I compute all of the Keplerian orbital quantities directly and then substitute them into the equations of \citetalias{Torres_1999}.

\begin{figure}
    \centering\includegraphics[width=\linewidth]{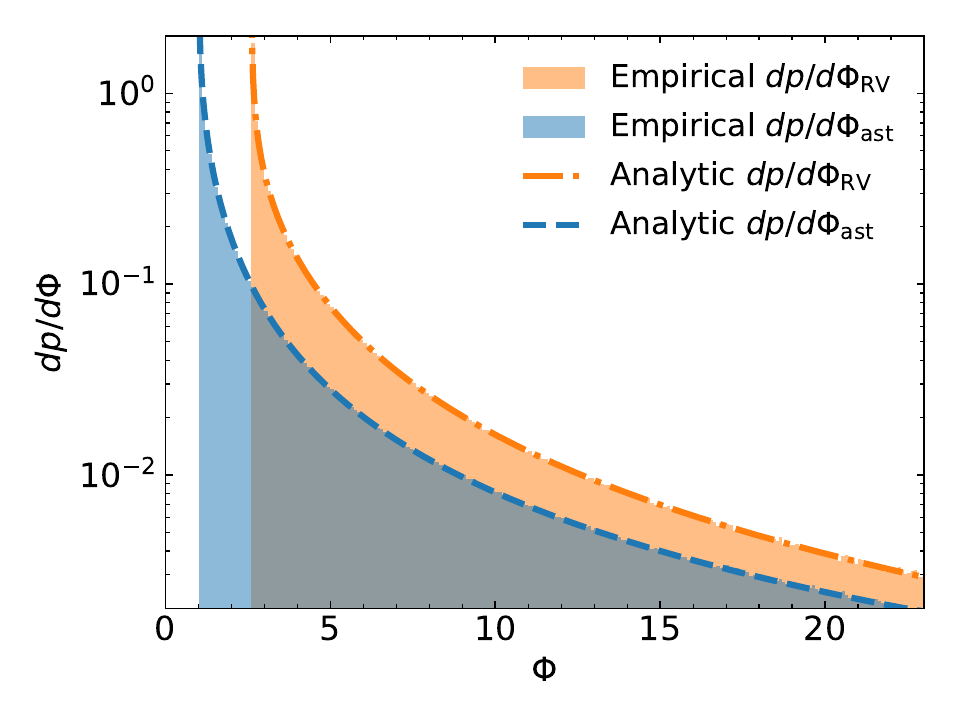}
    \caption{Empirical comparison of the probability density functions for $\Phi_{\rm RV}$ and $\Phi_{\rm ast}$ (Equations \eqref{eq:Phi_RV_final} and \eqref{eq:Phi_ast_dist}) to the equivalent formulae from \citetalias{Torres_1999} that are written as functions of the Keplerian orbital elements.  The empirical functions are histograms using $10^7$ points generated assuming random phases and orientations and a uniform distribution of eccentricities up to 0.8; they are indistinguishable from the analytic results.  \label{fig:verification}}
\end{figure}

Figure \ref{fig:verification} shows the results.  I empirically approximate the probability density functions using histograms and compare to the analytic results.  With $10^7$ random points, the empirical histograms and analytic distributions are indistinguishable.  I conclude that the expressions in Equations \eqref{eq:Phi_ast_dist} and \eqref{eq:Phi_RV_final} are indeed equivalent to the equations in \cite{Torres_1999}, so long as geometric orientations are random.  I note that Figure 2 of \citetalias{Torres_1999} looks very different from my result, either from the analytic expression or from applying the equations of \citetalias{Torres_1999} directly.  In particular, I am unable to reproduce their tail of inferred masses down to $M_2=0$ for a nonzero astrometric acceleration and projected separation.

\section{Ratio of Projected Separation to Semimajor Axis} 
\label{sec:projsep_semimajoraxis}

I next address the probability distribution of the ratio of the projected separation to the semimajor axis.  This probability distribution depends on the distribution of orbital eccentricities and is commonly calculated by Monte Carlo \citep{Torres_1999,Brandeker+Jayawardhana+Khavari+etal_2006,Dupuy+Liu_2011}.  I begin with the probability distribution of the true separation divided by the semimajor axis,
\begin{align}
    y \equiv \frac{r}{a} = 1 - e \cos E
\end{align}
where $e$ is the eccentricity and $E$ is the eccentric anomaly.  To obtain a probability distribution of $y$, I use the chain rule and Kepler's equation to write
\begin{align}
    \frac{dp}{dy} = \frac{1}{2\pi} \frac{dp}{dE} \left( \frac{dE}{d\eta} \right)
\end{align}
where $\eta \in U(0, 2\pi)$ is the orbital phase.  I have, after substitutions,
\begin{align}
    \frac{dp}{dy} = \frac{y}{\sqrt{e^2 - \left(y - 1\right)^2}}.
\end{align}

Next, I need to account for the unknown geometric orientation to convert $y$ into the ratio of the projected separation to semimajor axis; I will designate this ratio by $\psi$.  I have
\begin{equation}
    \psi = y \sin \varphi
\end{equation}
in direct analogy to the astrometric case in Section \ref{sec:analytic}.  I only consider positive values of $\sin \varphi$ since I am not distinguishing the cases where the companion is in front of or behind its host.  The domain of $y$ runs from $1-e$ to $1+e$, and $\psi$ can take on values from zero to $1+e$.

To obtain the probability distribution of $\psi$, I again apply the formula for the probability distribution of a product, Equation \eqref{eq:conditional_probability}.  I have
\begin{align}
    \frac{dp}{d\psi} &= \int \left( \frac{y}{\sqrt{e^2 - (y - 1)^2}} \right) \left( \frac{dy}{y} \right) \left( \frac{\psi/y}{\pi \sqrt{1 - \psi^2/y^2}}\right) \\
    &= \int \frac{1}{\pi} \frac{\psi dy}{\sqrt{\left(e^2 - (y - 1)^2\right)\left( y^2 - \psi^2\right)}} 
    \label{eq:integral_eccdist}
\end{align}
where the factor of $\pi$ normalizes $dp/d\varphi$.  The limits of integration depend on the value of $\psi$.  If $\psi < 1 - e$, then the integral runs from $y=1-e$ to $y=1+e$.  If $1 - e < \psi < 1 + e$, then the lower limit of integration is $\psi$.  If $\psi > 1 + e$, then the probability density is zero (this case is unphysical).  Equation \eqref{eq:integral_eccdist} is similar to Equation (43) of \cite{Savransky+Cady+Kasdin_2011}, but without those authors' integrals over eccentricity and semimajor axis.

Equation \eqref{eq:integral_eccdist} describes an elliptic integral.  For convenience, I define
\begin{equation}
    \alpha = \frac{4 e \psi}{1 - \left(e - \psi\right)^2} .
\end{equation}
The solution for $\psi < 1 - e$, i.e.~$\alpha < 1$, is
\begin{align}
    \frac{dp}{d\psi}\bigg|_{\psi < 1-e} = \sqrt{ \frac{\alpha \psi}{\pi^2 e}} K(\alpha) 
    \label{eq:pdist_1}
\end{align}
and the solution for $1 - e < \psi < 1 + e$, i.e.~$\alpha > 1$, is
\begin{align}
     \frac{dp}{d\psi}\bigg|_{|\psi - 1| < e} = \sqrt{ \frac{\psi}{\pi^2 e}} K(1/\alpha) 
     \label{eq:pdist_2}
\end{align}
where $K$ is the complete elliptic integral of the first kind.  $dp/d\psi$ diverges at $\psi = 1-e$, which presents a modest difficulty when integrating $dp/d\psi$ over $\psi$ or over $e$ (the divergence is integrable).  Equation \eqref{eq:pdist_1} is not valid for $e=0$ as written, but I can take the limit to obtain
\begin{equation}
    \frac{dp}{d\psi}\bigg|_{e=0} = \frac{\psi}{\sqrt{1-\psi^2}} .
\end{equation}

Equations \eqref{eq:pdist_1} and \eqref{eq:pdist_2} give the probability density $\partial p/\partial \psi$ at fixed $e$.  This probability density is normalized, with 
\begin{equation}
    \int_0^{1+e} \left( \frac{\partial p}{\partial \psi} \right) \bigg|_e d\psi = 1.
\end{equation}

To obtain $dp/d\psi$ for a distribution of eccentricities, I need to calculate
\begin{align}
    \frac{dp}{d\psi} = \int_0^1 \left(\frac{\partial p}{\partial \psi}\right)\bigg|_e p(e)\,de .
    \label{eq:pdist_integrated}
\end{align}
This integral must, in general, be computed numerically.  
\begin{figure*}
    \includegraphics[width=0.5\textwidth]{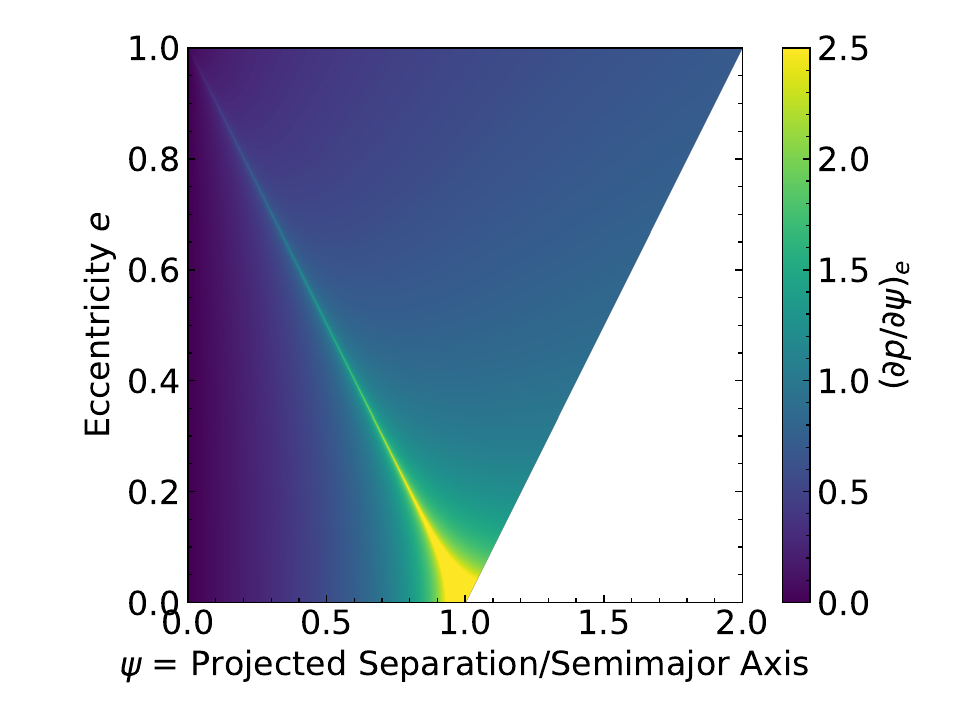}
    \includegraphics[width=0.5\textwidth]{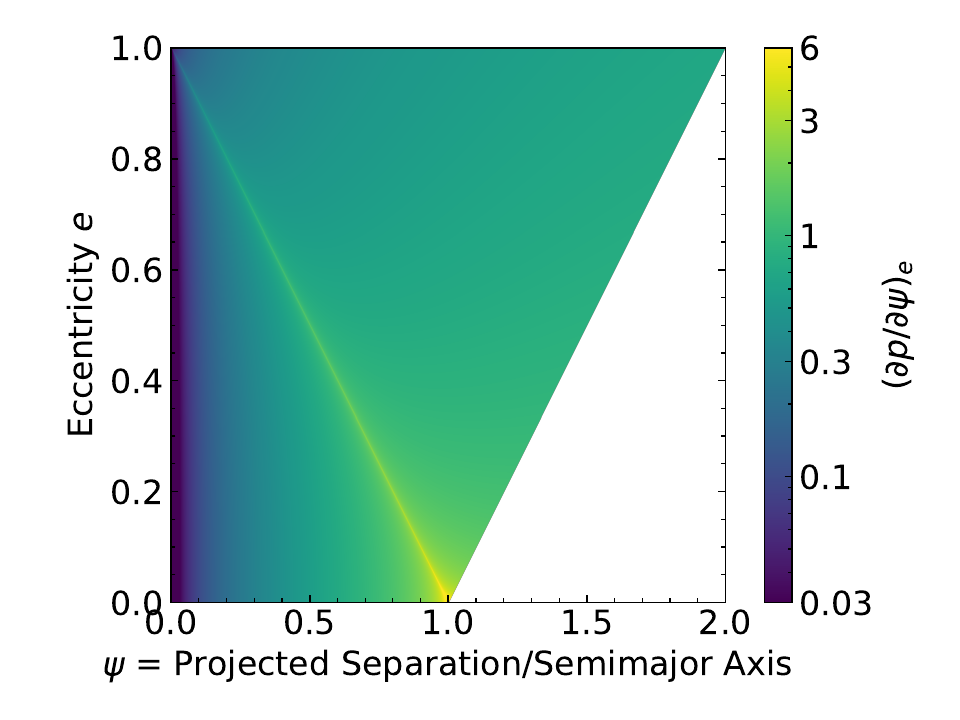}
    \caption{Two-dimensional probability density $d^2p/(d\psi\,de)$ as a function of eccentricity $e$ and $\psi$, the ratio of projected separation to semimajor axis.  The probability densities are calculated using Equations \eqref{eq:pdist_1} and \eqref{eq:pdist_2}, and are shown using linear (left) and logarithmic (right) color scales.  The region $\psi > 1 + e$ is forbidden on physical grounds, while there is an integrable singularity along the line $\psi = 1-e$. \label{fig:2D_maps}}
\end{figure*}
Figure \ref{fig:2D_maps} shows the probability density given in Equations \eqref{eq:pdist_1} and \eqref{eq:pdist_2} as a function of both eccentricity and $\psi$ using both linear (left) and logarithmic (right) color scales.  

\begin{figure*}
    \centering\includegraphics[width=0.495\linewidth]{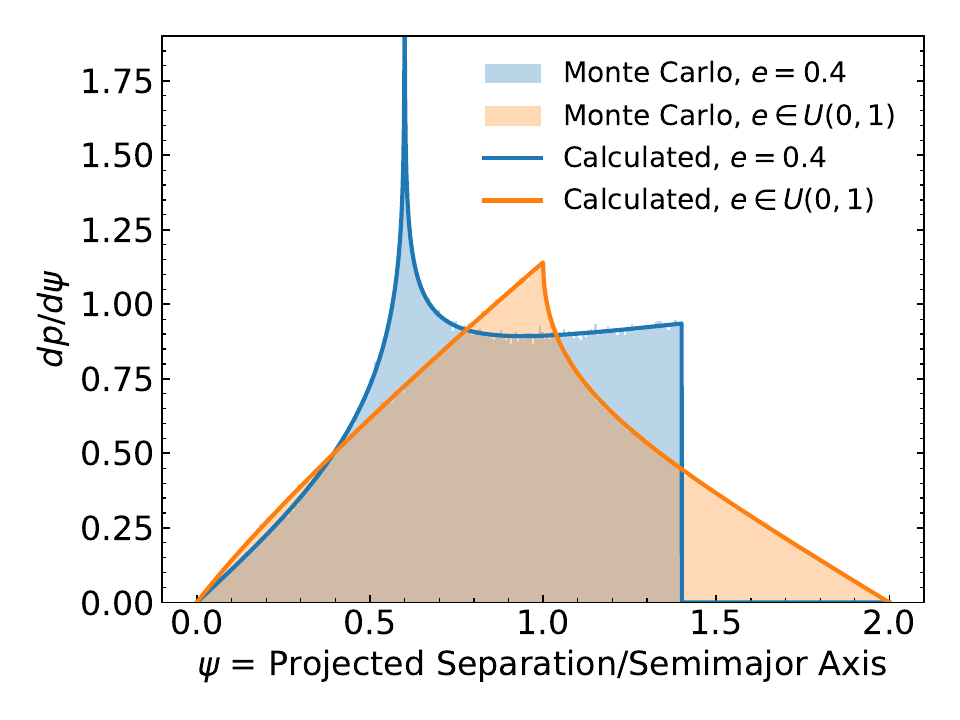}
    \includegraphics[width=0.495\linewidth]{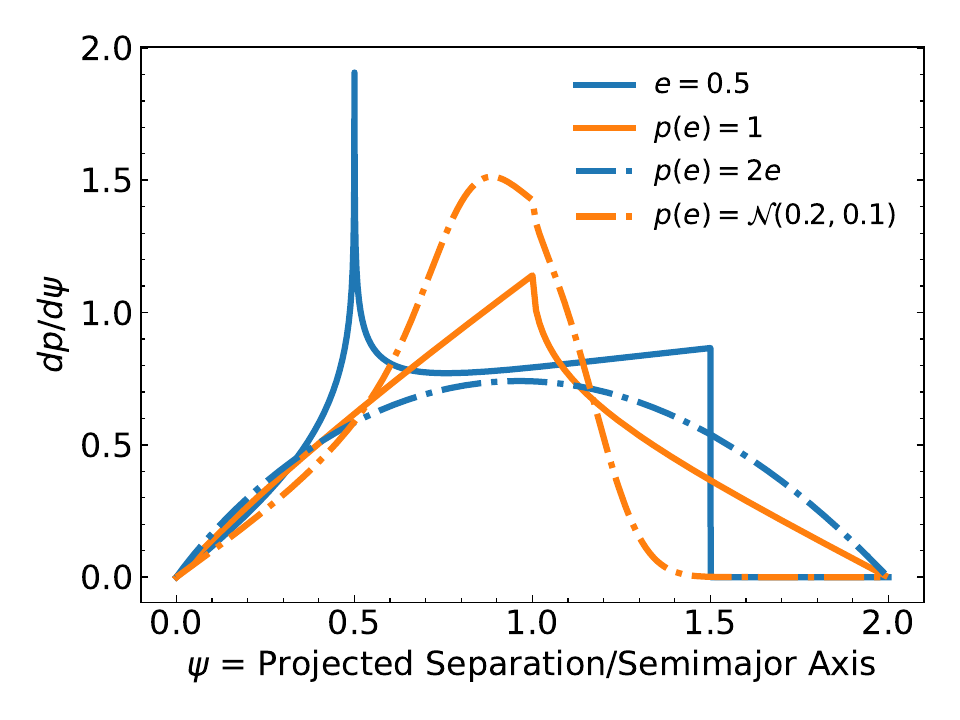}
    \caption{One-dimensional distributions of $\psi$, the ratio of projected separation to semimajor axis, for several distributions of eccentricity.  The left panel includes shaded histograms that have been computed from the equations of a Keplerian orbit using Monte Carlo.  Solid lines are computed using Equations \eqref{eq:pdist_1}, \eqref{eq:pdist_2}, and \eqref{eq:pdist_integrated}, 
    with the latter using numerical quadrature in one dimension. \label{fig:pdist_projsep_ecc}}
\end{figure*}

Numerically verifying the full maps shown in Figure \ref{fig:2D_maps} would be computationally expensive.  Instead, I perform two checks to verify the probability distribution of $\psi$: at fixed eccentricity, and under a uniform distribution of eccentricities.  The left panel of Figure \ref{fig:pdist_projsep_ecc} shows the results, with the shaded histograms using $10^6$ points to compute the probability distributions by Monte Carlo.  The solid lines are computed using Equations \eqref{eq:pdist_1} and \eqref{eq:pdist_2} (for the single eccentricity) or using numerical quadrature (for the uniform distribution of eccentricities).  

The right panel of Figure \ref{fig:pdist_projsep_ecc} shows the probability distributions of $\psi$ for a few additional distributions of eccentricities.  I use ${\cal N}(\mu, \sigma)$ to denote the normal distribution with mean $\mu$ and variance $\sigma^2$, truncated at $e=0$ and $e=1$ and renormalized over this domain.  In all cases apart from the single eccentricity, the probability distributions of $\psi$ are computed by one-dimensional numerical quadrature using {\tt numpy.integrate.quad}.  Future work could optimize the numerical integration using the properties of the complete elliptic integral.

\section{Discussion and Conclusions} \label{sec:conclusions}

In this paper, I have derived analytic probability density functions and cumulative distribution functions for the dimensionless ratio of a companion's true mass to the mass that would be inferred using a measured acceleration of the host star, either astrometric or radial, and a measured projected separation.  These functions may be used to statistically relate measured accelerations of stars with very long-period companions to the projected offsets of those companions from their host stars.  I have demonstrated the equivalence of these analytic expressions to probability distributions computed by Monte Carlo from the equations of Keplerian orbits.  The probability distributions of the dimensionless mass ratio given an RV or astrometric acceleration measurement are independent of all Keplerian orbital elements so long as the orientations of orbits are random.  

I have also derived closed-form solutions for the probability distribution of the ratio of the projected separation to the semimajor axis.  These expressions at fixed eccentricity may be written in terms of the complete elliptic integral of the first kind.  They may then be numerically integrated over an arbitrary distribution of eccentricities.  At fixed eccentricity, the probability distribution of this ratio has an integrable singularity in the interior of the domain.  Future work could implement efficient approaches to the numerical quadrature.

The expressions in this paper are implemented in an example Jupyter Python notebook\footnote{\url{https://github.com/
t-brandt/analytic-orbit-calculations}}, together with sufficient code to regenerate all figures.  The code in that notebook can serve as a template to implement the expressions for single calculations or within loops.

\noindent {\it Software}: scipy \citep{2020SciPy-NMeth},
          numpy \citep{numpy1, numpy2},
          matplotlib \citep{matplotlib},
          Jupyter (\url{https://jupyter.org/}).

\acknowledgements{I thank Dmitry Savransky for helpful feedback on this manuscript, and the organizers and attendees at the 25th annual NASA Sagan Summer School for providing much of the inspiration for this work.  I thank an anonymous referee for corrections and helpful suggestions, and in particular for suggesting the approach ultimately used to derive the cumulative distribution functions in Section \ref{sec:analytic}.}
          
\bibliography{refs}{}
\bibliographystyle{aasjournal}

\end{document}